\documentclass[conference]{IEEEtran}

\usepackage[utf8]{inputenc}
\usepackage{acronym}
\usepackage{multirow}
\usepackage[]{todonotes}
\usepackage{soul}
\usepackage{hyperref}
\usepackage[ruled,vlined]{algorithm2e}
\usepackage{listings}
\usepackage{array}
\usepackage{tabularx}
\usepackage{svg}
\usepackage{amsmath,amssymb,amsfonts}
\usepackage{pifont}
\usepackage[normalem]{ulem}
\usepackage{flushend}

\newcommand{\accessdate}{Accessed on 8 May 2024}

\usepackage[
    backend=biber,
    sorting=none,
    citestyle=numeric,
    bibstyle=numeric,
    maxbibnames=99,
]{biblatex}
\title{Scalable Data Notarization Leveraging Hybrid DLTs}

 \author{
 \IEEEauthorblockN{Domenico Tortola}
     \IEEEauthorblockA{
         \textit{University of Pisa}\\
         domenico.tortola@phd.unipi.it
     }
     \and
     \IEEEauthorblockN{Claudio Felicioli}
     \IEEEauthorblockA{
         \textit{Traent}\\
         claudio.felicioli@traent.com
     }
     \and
     \IEEEauthorblockN{Andrea Canciani}
     \IEEEauthorblockA{
         \textit{Geckosoft}\\
         a.canciani@geckosoft.it
     }
     \and
     \IEEEauthorblockN{Fabio Severino}
     \IEEEauthorblockA{
         \textit{Traent}\\
         fabio.severino@traent.com
     }
 }

\begin{document}

\maketitle


\begin{abstract}
    Notarization is a procedure that enhance data management by ensuring the authentication of data during audits, thereby increasing trust in the audited data. Blockchain is frequently used as a secure, immutable, and transparent storage, contributing to make data notarization procedures more effective and trustable. Several blockchain-based data notarization protocols have been proposed in literature and commercial solutions. However, these implementations, whether on public or private blockchains, face inherent challenges: high fees on public blockchains and trust issues on private platforms, limiting the adoption of blockchains for data notarization or forcing several trade-offs. In this paper, we explore the use of hybrid blockchain architectures for data notarization, with a focus on scalability issues. Through the analysis of a real-world use case, the data notarization of product passports in supply chains, we propose a novel approach utilizing a data structure designed to efficiently manage the trade-offs in terms of storage occupation and costs involved in notarizing a large collection of data.
\end{abstract}


\section{Introduction} \label{sec:introduction}

\noindent Recent years have witnessed significant interest in distributed ledger technologies (DLTs), with particular focus on blockchains. DLTs allow to share a database across a decentralized peer-to-peer network, ensuring that shared data remains immutable over time and transparently accessible to network members. These features have not only enabled the emergence of applications such as cryptocurrency trading, but have also enhanced other applications requiring improved security, transparency, and data immutability, such as data notarization. Notarizing data on blockchains in order to make them immutable and transparent is becoming a game changer feature in several real-world scenarios, such as the medical data sharing and IoT sensor data gathering, to cite the most relevant and more active in terms of research.

Data notarization has been effectively implemented on both public and private blockchains, each offering distinct advantages and drawbacks. Public blockchains ensure a high level of security, though they may have potentially high fees when writing data. Conversely, private blockchains have lower costs, but they lack the guarantees of data authenticity needed for external audits. Hybrid DLT architectures can overcome the drawbacks of notarizing data on public or private blockchains, providing an effective solution.

This paper assesses the scalability of a data notarization procedure implemented on a hybrid DLT architecture. The procedure meets several critical requirements, such as tamper evidence, historical consistency of notarized data, privacy preservation, cost efficiency and compactness for the data to be written to the public DLT. To efficiently handle the notarization of a large collection of data, we propose a novel approach that is able to satisfy the same requirements. We achieve the desired scalability by leveraging an authenticated and persistent novel data structure based on a bitwise trie.

The remainder of this paper is structured as follows: Section \ref{sec:background} introduces some background concepts, while Section \ref{sec:problem} describes a real-world use case where the data notarization procedure discussed in this paper is implemented, and gives a clear problem statement. In Section \ref{sec:solution} we describe the data structures that we propose to address the discussed problem, with some experimental evaluation about the size scalability presented in Section \ref{sec:experiments}. Section \ref{sec:conclusions} concludes the paper.

\section{Background} \label{sec:background}

\begin{figure*}[t]
    \centering
    \includegraphics[width=0.7\textwidth]{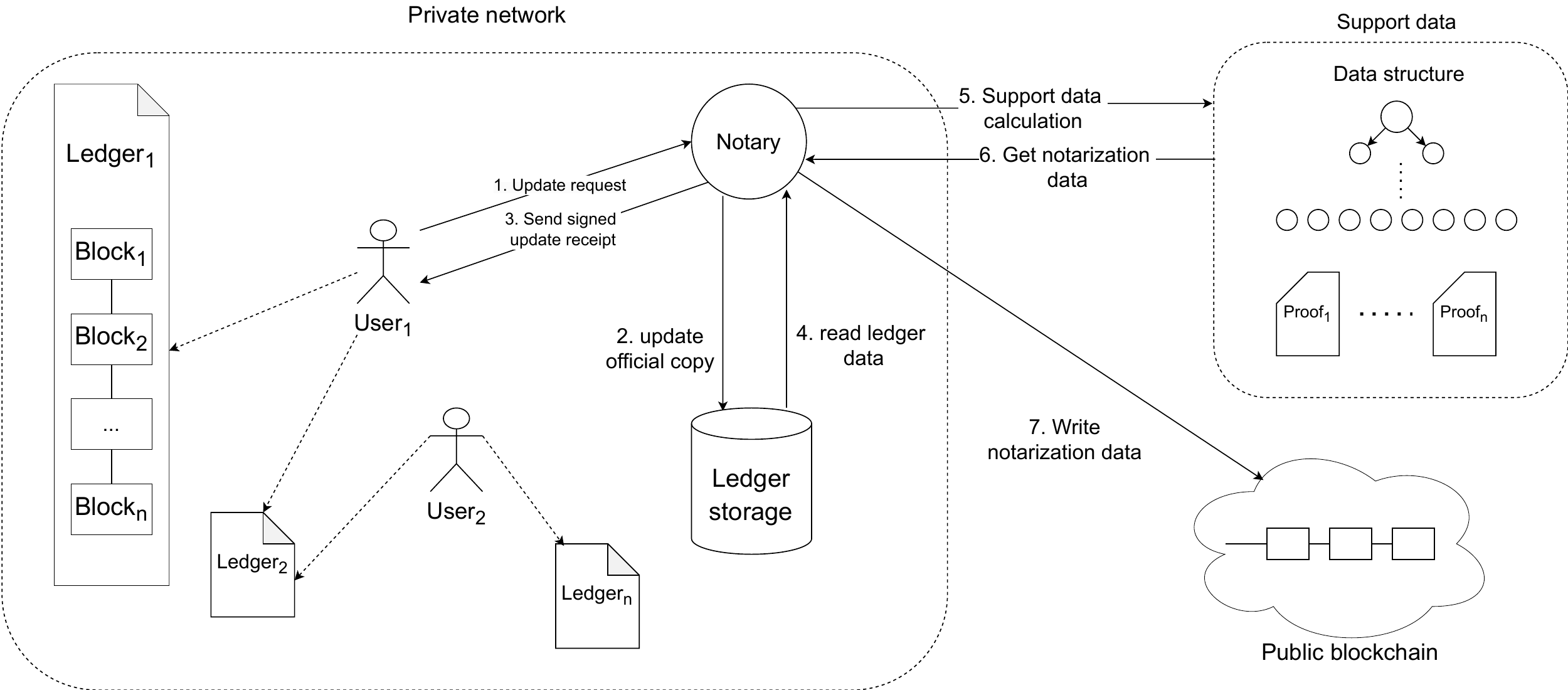}
    \caption{Hybrid DLT notarization framework overview}
    \label{fig:framework}
\end{figure*}

\subsection{Distributed Ledger Technologies} \label{sec:dlt}

\noindent Distributed ledger technologies (DLTs) defines a class of technologies where a database, or more generically a data structure, is shared across a network, aiming to achieve transparency, security and decentralization. Users in the network hold a copy of the data structure, which is synchronized after every update. As analyzed in \cite{IoiniP18dlt}, there are several implementations of DLTs (acyclic graphs, hashgraphs) but blockchains are the most popular. Popularized by Bitcoin and Ethereum, blockchains maintain a transaction ledger composed of data blocks. Each block contains transaction data and is linked to the previous one by a hash pointer. The blockchain is managed in a peer-to-peer network, where users hold a full copy of the ledger and update it every time a new block is created. According to how new members can join the network, blockchains are usually classified in two main categories: \textit{permissionless} (or public) and \textit{permissioned} (or private). Public blockchains, such as the mentioned Bitcoin and Ethereum, allow unrestricted access to join the network and provide a high level of transparency and user anonymity. In contrast, private blockchains like Hyperledger Fabric\footnote{\url{https://www.hyperledger.org/projects/fabric} [\accessdate]} restrict network access, and are a better choice to meet confidentiality requirements. Latest research lines propose an hybrid approach \cite{hybridDlt, PelosiFCS23ml}, which combines aspects from both public and privates DLTs.

\subsection{Data notarization} \label{sec:notarization}

\noindent Providing transparency and data immutability, blockchains are a natural support to notarize data. Considering that the main goal of data notarization is to provide an authenticated and tamper-proof data record keeping, the properties provided by a blockchain naturally fits such requirements. Notarization on public blockchain (e.g., Ethereum) is generally more secure and transparent, with an higher cost. On the other hand, the use of a private blockchain (e.g., Hyperledger Fabric) can reduce costs but but offers fewer guarantees of tamper-proofness (an external auditor can not have guarantees that audited data were not tampered with). Hybrid DLTs, combining elements from both public and private blockchains, can make a trade-off between costs and security, as explained in next sections.

Blockchain-based data notarization is a very promising research area with applications across various disciplines, including data sharing \cite{ChowdhuryCK0S18dataSharing}, biomedical \cite{KLEINAKI2018288biomedical}, smart services \cite{CossuLUF22smartServices}, and many others. 

\section{Problem definition} \label{sec:problem}

\subsection{A real world use case}

\noindent To have a full comprehension about the problem we want to address in this paper, it is useful to discuss a practical use case of hybrid DLT-based data notarization. We consider the use case analyzed in \cite{canciani24dpp}, which discuss blockchain-based product tracking through digital product passports (DPP) in supply chains. The paper emphasizes how an hybrid blockchain can be a reliable technological base to enforce the trust and the transparency in DPP and solve the problem of external auditability that affects implementations based on private DLTs: considering that in a private DLT members are known each other, it is not difficult to manipulate the consensus in order to create tampered data to deceive an external auditor.

The paper proposes to implement DPPs within an hybrid blockchain architecture using specific lightweight, portable data structures called \textit{ledgers}, a concatenation of data blocks handled in a private environment and notarized on a public blockchain, as discussed in \cite{hybridDlt}. This solution allows supply chain members to maintain the DPP and enable auditing it from both internal members and external entities (e.g., customers and authorities). To meet the requirements of this use case, the DPP should be auditable but also privacy-preserving, supporting a partial and selective disclosure of the data contained in the ledger according to the auditor needs.

Figure \ref{fig:framework} illustrates the notarization procedure. The figure highlights the components of the architecture and how they interact during the notarization process. In the private network, users produce and maintain ledger data structures. Every time a ledger changes, a signed update request is sent by the proposing user to the Notary module (step 1). The Notary maintains a dedicated storage, which contains a copy of each ledger, and processes the request updating the corresponding copy of the ledger in its storage (step 2). If the update is successful, a signed receipt is sent back to the user (step 3), where a dedicated mechanism disseminates the receipt over the private network in order to notify the ledger change to the other users. The notarization is executed periodically (in our case, once every 24 hours), with the Notary that reads the ledger data from its storage (step 4) and uses it to calculate the notarization support data (step 5), which may include a data structure and one or more proofs assessing the data structure consistency with its previous versions. Finally, the Notary calculates the notarization data, which consist of the data structure digest and eventually a consistency proof (step 6), and publishes them on a public blockchain (step 7). In our case, we choose Algorand\footnote{\url{https://developer.algorand.org/} [\accessdate]} as the public blockchain.

The simplest case is the notarization of a single ledger, i.e., a single DPP. In this case, the data structure calculated in the step 5 is a simple Merkle tree \cite{merkleTreeSurvey}, in which the data blocks of the ledger are treated as the leaf nodes of the tree (using the hash of the block as values) and internal nodes calculated according to the classical Merkle tree construction. The Merkle tree digest is written on a public blockchain, together with a Merkle consistency proof between such new digest and the previous one in the sequence. The intuition behind the consistency proof mechanism is that, if two chronologically ordered instances of a tree are consistent between each other, then the leaves of the older one are present, in the same order, as a prefix of the leaves of the newer tree.

\subsection{Problem statement}

\noindent The notarization of a single ledger is able to achieve external auditability supporting selective disclosure (the possibility to disclose only a portion of the ledger data), tamper-evidence (it is possible to verify the absence of tampering in the disclosed data by verifying the related proofs), consistency (the sequence of digests identify a provably consistent data history) and privacy preservation (the notarization process does not compromise the private network data secrecy, because only cryptographical digest and privacy-preserving proofs are published) and compactness (both the digest and the consistency proof can be stored in the \textit{note} field of an Algorand transaction, which is 1 kB). On the other hand, this solution does not scale adequately when the quantity of ledgers to notarize grows, because at each notarization the number of transactions on the public blockchain is linearly proportional to the number of ledgers to notarize. To give just an indicative cost analysis, a transaction on Algorand costs 0,001 Algos, and an Algo is currently valued \$0,19\footnote{\url{https://explorer.perawallet.app}, [\accessdate]}. The annual notarization cost, considering a single daily transaction as previously described, would be around \$0,07. Recalling the use case of the digital product passport, in which a single DPP is made by one or more ledgers, this cost would be acceptable when managing a moderately low quantity of high value products, but considering a production of 1.000.000 low value products would imply a minimum annual notarization costs around \$7.000 (considering just one ledger per DPP) which could not be economically viable, and, more alarmingly, the number of daily transactions required would surpass the current daily number of total transactions recorded in the Algorand network.

Therefore, the problem we aim to address is achieving the same properties as single ledger notarization (compactness, tamper-evidence, consistency, privacy preservation, and selective disclosure) while managing a large number of ledgers, but avoiding the linear proportionality between that number and the quantity of public blockchain transactions.

\section{Proposed solution} \label{sec:solution}

\begin{figure}
    \centering
    \includegraphics[width=0.46\textwidth]{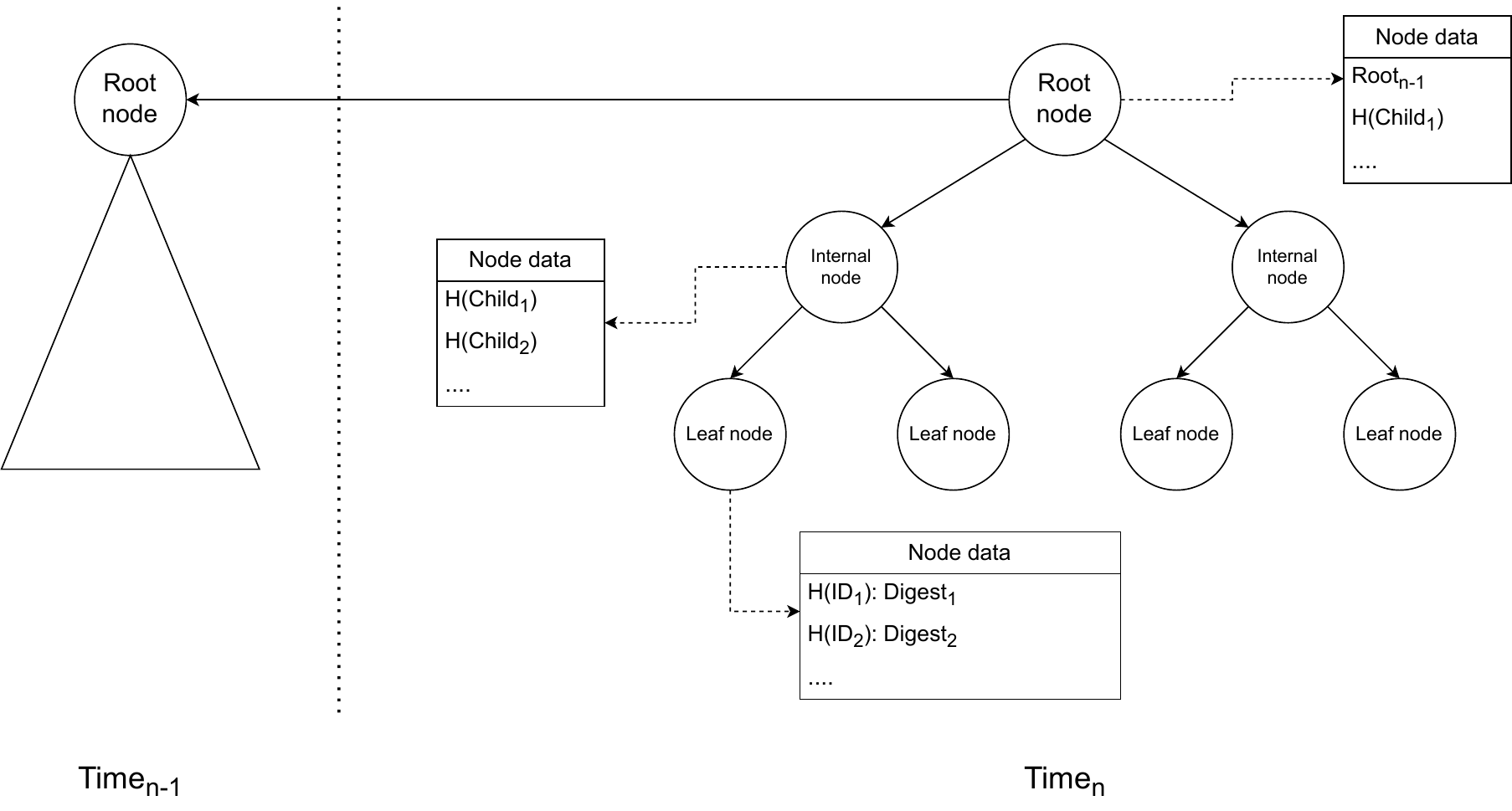}
    \caption{Data structure overview}
    \label{fig:data-structure}
\end{figure}

\noindent Our solution to address the scalability challenges of the notarization procedure involves replacing the Merkle tree used in step 5 of our single-ledger notarization procedure with a more advanced data structure. This novel data structure is designed to aggregate a potentially vast collection of authenticated and persistent child data structures. In the specific case under consideration, these child data structures are ledgers within a Hybrid DLT. For reference, Figure \ref{fig:data-structure} provides a high-level diagram of the data structure.

In Section \ref{sec:construction}, we illustrate the basic construction of the data structure. Sections \ref{sec:authentication} and \ref{sec:persistency} delve into the key properties of this data structure and how they are achieved. In Section \ref{sec:proof} we illustrate how it is possible to rely only on the historical sequence of digests of this data structure to audit the consistency of each individual child data structure.

By notarizing just a single sequence of digests, we are then able to achieve external auditability for all the child data structures. This is a more scalable approach than the single-ledger solution discussed in Section \ref{sec:problem}, where notarization of a different sequences of digest for each individual ledger was required. Furthermore, being able to notarize an arbitrary number of ledgers using a single compact digest means that the costs due to public blockchain writes remains constant, becoming effectively negligible.

\subsection{Construction} \label{sec:construction}

\noindent The structure we propose is based on an $r$-ary bitwise trie, which facilitates the implementation of a dictionary. In our case, the search keys are the binary representations of the hashes of child data structure identifiers. The values associated with these keys are the digests of the child data structures themselves.

The trie used in our construction has two kinds of nodes: internal or leaf. Each internal node can have up to $r$ children, with $r$ being a power of $2$. The outgoing edges are represented within the node through a list of cells of fixed length $r$, where each position in the list is associated a priori with one of the possible edge labels (the $r$ configurations of $log_2(r)$ bits). Each cell contains either a reference to a child node or the special value $null$. Leaves contain a set of tuples $<key, value>$, with a maximum cardinality of $k$. In order to be valid, each $key$'s binary representation must be prefixed by the concatenation of the edge labels of the path from the root to the leaf (the search path) that contains the tuple.


\subsection{Authentication} \label{sec:authentication}

\noindent The data structure is authenticated, with a cryptographic digest identifying its state. In order to describe the digest of the data structure, we first define the digests of the nodes: we establish a canonical serialization for them and then we use this serialization as the input for a cryptographic hash function, such as SHA-512. The resulting hash value serves as the node’s digest. These digests are utilized in the internal nodes as the references to child nodes.

The hash of the root node is used as the digest that identifies the entire trie’s state. Indeed, similar to what occurs in Merkle trees, any change to the state of any node not only alters the node’s digest but also affects its parent node, as it contains the hash value of the child. This triggers a cascade of digest modifications along the entire path, up to and including the root.

\subsection{Persistency} \label{sec:persistency}

Adopting the terminology introduced in \cite{driscoll1986making}, an \textit{ephemeral} data structure is characterized by the loss of its previous state upon any modification. In contrast, a \textit{persistent} data structure allows for accessing various versions of its state, and the execution of an edit operation results in the the creation of a new state rather than replacing previously stored states. There are several forms of persistence, but the type we are particularly interested in for our specific case is \textit{partial persistence}. For a data structure to be partially persistent, it must save all its states within a version history, arranged according to the temporal evolution of the structure. Edit operations can only be performed on the most recent version, while query operations can be executed on any version in the history.

To obtain partial persistence in our proposed data structure, any modifications to the current state do not alter existing nodes. Instead, all the existing nodes states are preserved, while a new version of the trie is created. As the node references are hash values of nodes, unchanged portions of the trie are effectively reused in the new version. This method, broadly referred to as \textit{path-copying} in \cite{driscoll1986making}, is a technique used to achieve persistence in linked data structures.

To maintain the authenticity of the data structure, we introduce an additional modification: the root node is extended to include not only the digests of its children but also the digest of the trie's previous root version.

\subsection{Notarization procedure} \label{sec:proof}

The purpose of this data structure is to reduce the notarization data that must be recorded on the public blockchain to only the historical sequence of digests of this data structure. Here we describe a procedure that allows a privacy-preserving external audit for each child data structure, based exclusively on this sequence of digests.

For the formal definition of a privacy-preserving external audit, we refer to \cite{canciani2023auditable}. Here we assume that the child data structures support the construction of privacy-preserving consistency proofs (in our specific use case, the child data structures are ledgers, and, as described in Section \ref{sec:problem}, the privacy-preserving proof is the Merkle consistency proof).

For each specific child data structure with identifier $ID$, we aim to enable verification of the following properties:

\begin{itemize}
    \item No alternative histories for $ID$: a single value (either a digest or $null$) must be associated to $ID$ at every notarization.
    \item No removal for $ID$: once an association between $ID$ and a digest has been notarized, an association between $ID$ and a digest must also be present in all subsequent notarizations.
    \item No forks in $ID$’s history: the list of digests associated with $ID$ must form a history, consistent with the valid operations defined for the child data structure.
\end{itemize}

The notarization procedure is performed periodically (in our use case, once every 24 hours) and involves the following steps. First, a snapshot of the state of all child data structures to be notarized is taken. This state is represented as a compact digest. If a child data structure was also included in the previous notarization and its digest has changed in the current notarization, a privacy-preserving consistency proof between the two digests (in our use case, a Merkle consistency proof) is generated. Once a child data structure is added to the list of notarized data structures, it must always be included in all subsequent notarizations. From the list of $<ID, digest>$ associations, a new trie is constructed. The root of this new trie must includes a reference to the root of the trie generated in the previous notarization. The digest of the root of the new trie is then written to the public blockchain. Finally, all newly generated nodes and consistency proofs are published in a publicly accessible storage.

When data contained in a child data structure are shared outside the private network (even partially, as described in \cite{hybridDlt} and enabled by authenticated data structures), anyone can access the data stored in the public storage along with the sequence of digests recorded on the public blockchain to perform an audit, which verifies the previously described properties. The audit procedure consists of the following steps:

\begin{itemize}
    \item Identify the current root, i.e., the node whose digest matches the latest digest published on the public blockchain.
    \item Search within the trie using the hash of the child data structure's $ID$, verifying the correctness of the hash references in the internal nodes, and ultimately obtaining the value for the child data structure’s digest.
    \item Repeat the search for all previous versions of the trie (they can be found thanks to the link to the preceding trie version contained in the root nodes), thus compiling a list of digest values for the child data structure.
    \item Verify that the list of digests of the traversed roots matches the list of digests published on the public blockchain.
    \item Verify that the list of digest values for the child data structure, previously derived from the searches, are all non-$null$, with the only eventual exception of a prefix of $null$ values.
    \item For every change in value along the list of digest for the child data structure, retrieve the associated consistency proof from the public access storage and verify it.
    \item Verify that the digest of the shared child data structure matches one of the values associated with it in the notarizations, or alternatively, verify that consistency proofs have been shared which confirm that the digest is consistent with one of the list's values, and that the subsequent value is in turn consistent with the digest.
\end{itemize}

An alternative auditing procedure that does not require accessing the public storage involves constructing an audit proof. This audit proof simply comprises all the nodes and proofs that would normally be retrieved from the public storage during the execution of the previously described auditing procedure. An audit proof can be generated on-demand when a child data structure is intended to be shared.

\subsection{Limitations} \label{sec:limitations}


\noindent The main limitation is the size of the trie. It is unfeasible to store it directly in the public blockchain and therefore, as described in the previous section, the trie is published on a public access storage. As a possible trade-off for not relaying on the public access storage, is to generate audit proof when sharing the content of a child data structure.

A limitation of the method based on the audit proofs is the static nature of the proof, which allows validation only up to the last notarization event at the time of its creation. Nevertheless, the public access storage can still be used to verify the continued validity of the properties in subsequent notarizations, but limiting the advantage of the trade-off.

Another limit of audit proofs is that each of them can be used to verify the historical consistency of a single ledger. At the moment, we are not able to propose a consistency proof for the full trie that is compact.

\section{Size scalability evaluation} \label{sec:experiments}

\noindent To better evaluate the scalability of the proposed solution, we conducted multiple tests, adjusting the $k$ and $r$ parameters, to analyze the growth of key indicators as the number of notarized ledgers increases. The code of the experiments is publicly accessible on GitHub\footnote{\url{https://github.com/pangon/notary-data-structure-simulation} [\accessdate]}.

Given that one of the notarization procedures we propose involves publishing the data structure on a public access storage, one key indicators of interest is undoubtedly the size of the trie. To compute it, we consider an implementation where nodes are serialized efficiently. In this implementation, internal nodes use a bitmap of $r$ bits to mark every non-null branch, while the serialization of leaf nodes includes a header of $log_2(k)$ bits specifying the number of tuples contained.

Regarding the audit proof, its size depends on several factors: notarization frequency, the size of the search paths in the trie, and the size of the included consistency proofs. Of these factors, the only one influenced by the trie parameter choices is the size of the search paths. The notarization frequency is fixed and impacts the proof size linearly. The size of a single consistency proof depends on the specific type of child data structure used; for example, in the case of ledgers, it involves Merkle consistency proofs, whose size is logarithmic relative to the number of nodes in the ledger. The number of consistency proofs included depends on the frequency of state changes in the child data structure, and is entirely determined by the use case. For instance, a ledger used to track measurements from an IoT device might grow to millions of nodes, with its state changing with each notarization. Conversely, a ledger used for a product passport, primarily tied to documenting a production process over a limited period, might reach a maximum size of only a few hundred blocks and then remain largely inactive.

With this in mind, the most significant indicators influenced by the choice of the two parameters, $r$ (the maximum number of children) and $k$ (the maximum number of tuples in a leaf), are the total size of the trie produced during a notarization, and the average size of a search path in the trie.

An initial observation is that the trie is unbalanced. However, the use of hash values of the $ID$s as search keys helps mitigate the risk of extreme imbalance, which could arise from unfortunate ID choices or deliberate poisoning attacks. During experimentation, the search paths showed skewed distributions, but they consistently exhibited a logarithmic relationship with the number of notarized child data structures.

\begin{table*}[!t]
\renewcommand{\arraystretch}{1.3}
\caption{Measurements with different configurations of $r$ and $k$ parameters}
\label{tab:measurements}
\centering
\begin{tabular}{ccc||r|rrr|r|c}
\hline
\multicolumn{3}{c||}{\bfseries configuration} & \multicolumn{6}{c}{\bfseries measurements} \\
\hline
\bfseries r & \bfseries k & \bfseries ledgers count & \bfseries nodes count & \multicolumn{3}{c|}{\bfseries search path length} & \bfseries total size (bytes) & \bfseries search path \\
 & & & & minimum & maximum & average & & \bfseries average size (bytes) \\
\hline
\hline
2 & 1 & 10.000.000 & 24.428.458 & 21 & 49 & 25,59 & 2.207.028.362 & 3.217 \\
2 & 2 & 10.000.000 & 13.604.938 & 20 & 37 & 24,14 & 1.513.318.087 & 3.170 \\
2 & 4 & 10.000.000 & 7.144.399 & 20 & 29 & 22,94 & 1.099.027.571 & 3.217 \\
2 & 8 & 10.000.000 & 3.610.063 & 20 & 26 & 21,85 & 872.172.029 & 3.448 \\
4 & 1 & 10.000.000 & 17.225.679 & 11 & 24 & 13,54 & 1.746.056.231 & 3.153 \\
4 & 2 & 10.000.000 & 10.879.805 & 11 & 18 & 12,82 & 1.339.017.320 & 3.141 \\
4 & 4 & 10.000.000 & 6.369.408 & 11 & 15 & 12,21 & 1.049.673.604 & 3.185 \\
4 & 8 & 10.000.000 & 3.869.219 & 11 & 14 & 11,74 & 889.205.391 & 3.326 \\
8 & 1 & 10.000.000 & 14.727.206 & 8 & 16 & 9,52 & 1.587.268.326 & 4.052 \\
8 & 2 & 10.000.000 & 10.517.979 & 8 & 13 & 9,07 & 1.316.636.921 & 4.048 \\
8 & 4 & 10.000.000 & 7.353.018 & 8 & 11 & 8,70 & 1.113.475.705 & 4.080 \\
8 & 8 & 10.000.000 & 3.037.128 & 8 & 9 & 8,11 & 835.773.176 & 4.258 \\
\hline
\end{tabular}
\end{table*}



Table \ref{tab:measurements} presents some results from our experiments, which varied the parameters $r$ and $k$. We analyzed data structure instances constructed from $10.000.000$ ledgers, testing several combinations of $r = [2, 4, 8]$ and $k = [1, 2, 4, 8]$. For each instance, we measured the nodes count, the length of the search paths (shortest, longest, and average), the total size of the trie's serialization, and the average size of the serialized search paths.

We can observe that although an increase of the $r$ parameter consistently reduces the length of the search path, it does not necessarily decrease also the search path size. This occurs because the value of $r$ reduces the path length by a factor of its logarithm, but simultaneously increases the size of the intermediate nodes linearly.

Regarding the $k$ parameter, its variation impacts the total size of the structure but has a much lesser effect on the search path size. We observed that increasing $k$ maintains unchanged the cumulative size of the leaves, due to optimized leaf serialization. This $k$ increase results also in fewer internal nodes, thereby saving space; however, the leaf nodes become individually larger and are shared among more ledgers, increasing the size of the search paths. The actual purpose of the parameter $k$ is to reduce the skewness in the search path length distribution that arises from using an unbalanced trie. As $k$ increases, the average length of the search path approaches the expected value for a balanced tree, mitigating the extreme case often observed when $k$ is set to $1$, where the maximum search path lengths can be twice as long as the minimum.

\section{Conclusions} \label{sec:conclusions}

\noindent In this paper, we explored the use of hybrid DLT for data notarization, with a focus on scalability issues. We analysed a real-world use case, the data notarization of product passports in supply chains. We describe a simple notarization procedure for a single ledger, a solution that already achieves all the ideal properties of notarization. After discussing the scalability limitations of this first procedure, we propose more advanced notarization procedures based on a novel data structure. This data structure enables the efficient notarization of an arbitrary number of ledgers. After discussing the trade-offs associated with this new method, we presented the results of our tests using several combinations of the data structure parameters, simulating the notarization of millions of ledgers.

As future work, we plan to describe new protocols for using the audit proofs generated from the data structure we have presented. These protocols will involve a network of independent monitors to reduce the data that must be shared along with the ledgers when they are exported from the private network.

\printbibliography

@inproceedings{IoiniP18dlt,
  author       = {Nabil El Ioini and
                  Claus Pahl},
  title        = {A Review of Distributed Ledger Technologies},
  booktitle    = {On the Move to Meaningful Internet Systems. {OTM} 2018 Conferences
                  - Confederated International Conferences: CoopIS, C{\&}TC, and
                  {ODBASE} 2018},
  series       = {Lecture Notes in Computer Science},
  volume       = {11230},
  pages        = {277--288},
  publisher    = {Springer},
  year         = {2018}
}

@article{canciani2023auditable,
  author       = {Andrea Canciani and
                  Claudio Felicioli and
                  Fabio Severino and
                  Domenico Tortola},
  title        = {Auditable data structures: theory and applications},
  journal      = {CoRR},
  volume       = {abs/2306.01886},
  year         = {2023},
  url          = {https://doi.org/10.48550/arXiv.2306.01886},
  doi          = {10.48550/ARXIV.2306.01886},
  eprinttype    = {arXiv}
}

@article{hybridDlt,
  title={Hybrid DLT as a data layer for real-time, data-intensive applications},
  author={Canciani, Andrea and Felicioli, Claudio and Lisi, Andrea and Severino, Fabio},
  journal={arXiv preprint arXiv:2304.07165},
  year={2023}
}

@INPROCEEDINGS{merkleTreeSurvey,  
    author={Liu, Haojun and Luo, Xinbo and Liu, Hongrui and Xia, Xubo},  
    booktitle={2021 International Conference on Electronic Information Engineering and Computer Science (EIECS)},
    title={Merkle Tree: A Fundamental Component of Blockchains},   
    year={2021},  
    volume={},  
    number={},  
    pages={556-561},  
    doi={10.1109/EIECS53707.2021.9588047}
}

@INPROCEEDINGS{canciani24dpp,
  author={Canciani, Andrea and Felicioli, Claudio and Severino, Fabio and Tortola, Domenico},
  booktitle={2024 IEEE International Conference on Pervasive Computing and Communications Workshops and other Affiliated Events (PerCom Workshops)}, 
  title={Enhancing Supply Chain Transparency through Blockchain Product Passports}, 
  year={2024},
  doi={10.1109/PerComWorkshops59983.2024.10502429}
}

@inproceedings{ChowdhuryCK0S18dataSharing,
  author       = {Mohammad Jabed Morshed Chowdhury and
                  Alan Colman and
                  Muhammad Ashad Kabir and
                  Jun Han and
                  Paul Sarda},
  title        = {Blockchain as a Notarization Service for Data Sharing with Personal
                  Data Store},
  booktitle    = {17th {IEEE} International Conference On Trust, Security And Privacy
                  In Computing And Communications / 12th {IEEE} International Conference
                  On Big Data Science And Engineering, TrustCom/BigDataSE 2018},
  pages        = {1330--1335},
  publisher    = {{IEEE}},
  year         = {2018},
  url          = {https://doi.org/10.1109/TrustCom/BigDataSE.2018.00183},
}

@article{KLEINAKI2018288biomedical,
  title = {A Blockchain-Based Notarization Service for Biomedical Knowledge Retrieval},
  author = {Kleinaki, Athina-Styliani and Mytis-Gkometh, Petros and Drosatos, George and Efraimidis, Pavlos S and Kaldoudi, Eleni},
  journal = {Computational and Structural Biotechnology Journal},
  volume = {16},
  pages = {288-297},
  year = {2018},
  issn = {2001-0370},
  doi = {https://doi.org/10.1016/j.csbj.2018.08.002},
}

@inproceedings{CossuLUF22smartServices,
  author       = {Raimondo Cossu and
                  Maria Ilaria Lunesu and
                  Marco Uras and
                  Alessandro Floris},
  title        = {A Blockchain-based Data Notarization System for Smart Mobility Services},
  booktitle    = {{IEEE} International Conference on Software Analysis, Evolution and
                  Reengineering, {SANER} 2022, Honolulu, HI, USA, March 15-18, 2022},
  pages        = {1231--1238},
  publisher    = {{IEEE}},
  year         = {2022},
  url          = {https://doi.org/10.1109/SANER53432.2022.00146}
}

@inproceedings{driscoll1986making,
  title={Making data structures persistent},
  author={Driscoll, James R and Sarnak, Neil and Sleator, Daniel Dominic and Tarjan, Robert Endre},
  booktitle={Proceedings of the eighteenth annual ACM symposium on Theory of computing},
  pages={109--121},
  year={1986}
}

@inproceedings{PelosiFCS23ml,
  author       = {Andrea Pelosi and
                  Claudio Felicioli and
                  Andrea Canciani and
                  Fabio Severino},
  title        = {A Hybrid-DLT Based Trustworthy {AI} Framework},
  booktitle    = {{IEEE} International Conference on Enabling Technologies: Infrastructure
                  for Collaborative Enterprises, {WETICE} 2023},
  pages        = {1--6},
  publisher    = {{IEEE}},
  year         = {2023},
  doi          = {10.1109/WETICE57085.2023.10477792},
}

\end{document}